\documentclass[aps,showpacs,floatfix,twocolumn,byrevtex,superscriptaddress]{revtex4-1}

\usepackage{amsmath}
\usepackage{amssymb}
\usepackage{amstext}
\usepackage{amsopn}
\usepackage{amsfonts}
\usepackage{amsxtra}
\usepackage[english]{babel}
\usepackage{amsmath}
\usepackage{graphicx}
\usepackage{float}
\usepackage{bm}
\usepackage{multirow}
\usepackage{dcolumn}
\usepackage{hyperref}
\usepackage{CJK}

\begin{document}

\title{Temperature induced band shift in ferromagnetic Weyl semimetal Co$_{3}$Sn$_{2}$S$_{2}$ }
\author{Run Yang}
\thanks{These authors contributed equally to this work.}
\author{Tan Zhang}
\thanks{These authors contributed equally to this work.}
\author{Liqin Zhou}
\affiliation{Beijing National Laboratory for Condensed Matter Physics, Institute of Physics, Chinese Academy of Sciences, Beijing 100190, China}
\affiliation{School of Physical Sciences, University of Chinese Academy of Sciences, Beijing 100049, China}
\author{Yaomin Dai}
\affiliation{Center for Superconducting Physics and Materials, National Laboratory of Solid State Microstructures and Department of Physics, Nanjing University, Nanjing 210093, China}
\author{Zhiyu Liao}
\affiliation{Beijing National Laboratory for Condensed Matter Physics, Institute of Physics, Chinese Academy of Sciences, Beijing 100190, China}
\affiliation{School of Physical Sciences, University of Chinese Academy of Sciences, Beijing 100049, China}
\author{Hongming Weng}
\email{hmweng@iphy.ac.cn}
\affiliation{Beijing National Laboratory for Condensed Matter Physics, Institute of Physics, Chinese Academy of Sciences, Beijing 100190, China}
\affiliation{School of Physical Sciences, University of Chinese Academy of Sciences, Beijing 100049, China}
\affiliation{Songshan Lake Materials Laboratory, Dongguan, Guangdong 523808, China}
\affiliation{CAS Center for Excellence in Topological Quantum Computation, Beijing 100190, China}
\author{Xianggang Qiu}
\email{xgqiu@iphy.ac.cn}
\affiliation{Beijing National Laboratory for Condensed Matter Physics, Institute of Physics, Chinese Academy of Sciences, Beijing 100190, China}
\affiliation{School of Physical Sciences, University of Chinese Academy of Sciences, Beijing 100049, China}
\affiliation{Songshan Lake Materials Laboratory, Dongguan, Guangdong 523808, China}
\date{\today}
%
\begin{abstract}
The discovery of nonmagnetic Weyl semimetals (WSMs) in TaAs compounds has triggered lots of efforts in finding its magnetic counterpart. While the direct observation of the Weyl nodes and Fermi arcs in a magnetic candidate through angle-resolved photoemission spectroscopy is hindered by the complex magnetic domains. The transport features of magnetic WSMs, including negative magnetoresistivity and anomalous Hall conductivity, are not conclusive since these are sensitive to extrinsic factors like defects and disorders in lattice or magnetic ordering.
Here, we systematically study the temperature dependent optical spectra of ferromagnetic Co$_3$Sn$_2$S$_2$ experimentally and simulated by first-principles calculations. The many-body correlation effect due to Co $3d$ electrons leads to renormalization of bands by a factor about 1.33, which is moderate and the description within density functional theory is suitable. As temperature drops down, the magnetic phase transition happens and the magnetization drives the band shift through exchange splitting. The optical spectra can well detect these changes, including the transitions sensitive and insensitive to the magnetization, and those from the bands around the Weyl nodes. The results strongly support that Co$_3$Sn$_2$S$_2$ is a magnetic WSM and the Weyl nodes can be tuned by magnetization with temperature change.
\end{abstract}


\maketitle

%
%
Recently, the shandite compound Co$_{3}$Sn$_{2}$S$_{2}$ has attracted lots of attentions since it not only shows intrinsic ferromagnetism but also is proposed to have three pairs of Weyl points around the Fermi level in the first Brillouin zone (BZ)~\cite{Liu2018, Wang2018, Xu2018, 2019arXiv190300509M}.
For its quasi-two-dimensional crystal structure, low carrier density and strong anomalous Hall effect (AHE), Co$_{3}$Sn$_{2}$S$_{2}$ has been thought as a potential candidate to realize the quantum AHE~\cite{Yu2010, Weng2015a} in its tow-dimensional (2D) limit~\cite{Muechler2017, Liu2018,PhysRevLett.107.186806}.
As a candidate of magnetic Weyl semimetal (WSM) \cite{PhysRevB.83.205101}, the magnetic ordering states are expected to have interactions with Weyl nodes~\cite{Xu2018, Wang2018}, so that the topological properties from WSM can be finely tuned through control of magnetization~\cite{Zhang2017, Wang2018}.
Actually, it has been found that upon cooling, the magnetization and anomalous Hall conductivity (AHC) of Co$_{3}$Sn$_{2}$S$_{2}$ vary accordingly~\cite{Wang2018, Liu2018} and the first-principles calculations of AHC at different magnetization has shown their intrinsic relationship through the changes in the position of Weyl nodes~\cite{Wang2018}.
To check whether this picture is true or not in realistic samples, here we have performed systematical optical spectra measurements under different temperatures ($T$s).
The advantages of optical spectra measurements over the AHC measurements are as following: First, optical spectrum is sensitive to both occupied and unoccupied bands, which constitute the Weyl nodes in WSM. In contrast, AHC is only sensitive to the one crossing Fermi level~\cite{Nagaosa2010, Xiao2010, Weng2015a, Yue2017, Steiner2017, Fang92}. Second, on considering optical selection rule and peak position, one can definitely identify the contributions from Weyl nodes~\cite{Xu2016, Shao2019, Armitage2018}. Third, transport measurements like AHC requires high techniques and high quality of the whole device~\cite{Liang2018, Yue2017}. The extrinsic factors make the understanding of intrinsic mechanism very difficult.

In this work, combining optical spectroscopy and first-principle calculations, we systematically investigate the band structure of Co$_{3}$Sn$_{2}$S$_{2}$ at various $T$s above and below its Curie temperature ($T_{C}$).
We have clearly demonstrated how the band structure evolves and identified both the magnetization sensitive and nonsensitive contributions to different optical conductivity peaks.
The contributions around the Weyl nodes are also identified.
Furthermore, we have found that the electron-electron correlation effect in this magnetic compound is not strong and the band renormalization factor is around 1.33.
The first-principles calculations within local density functional approximation is suitable.
These give out strong supports that Co$_{3}$Sn$_{2}$S$_{2}$ is a magnetic WSM.

%
%
\begin{figure}[tb]
\centerline{
\includegraphics[width=1\columnwidth]{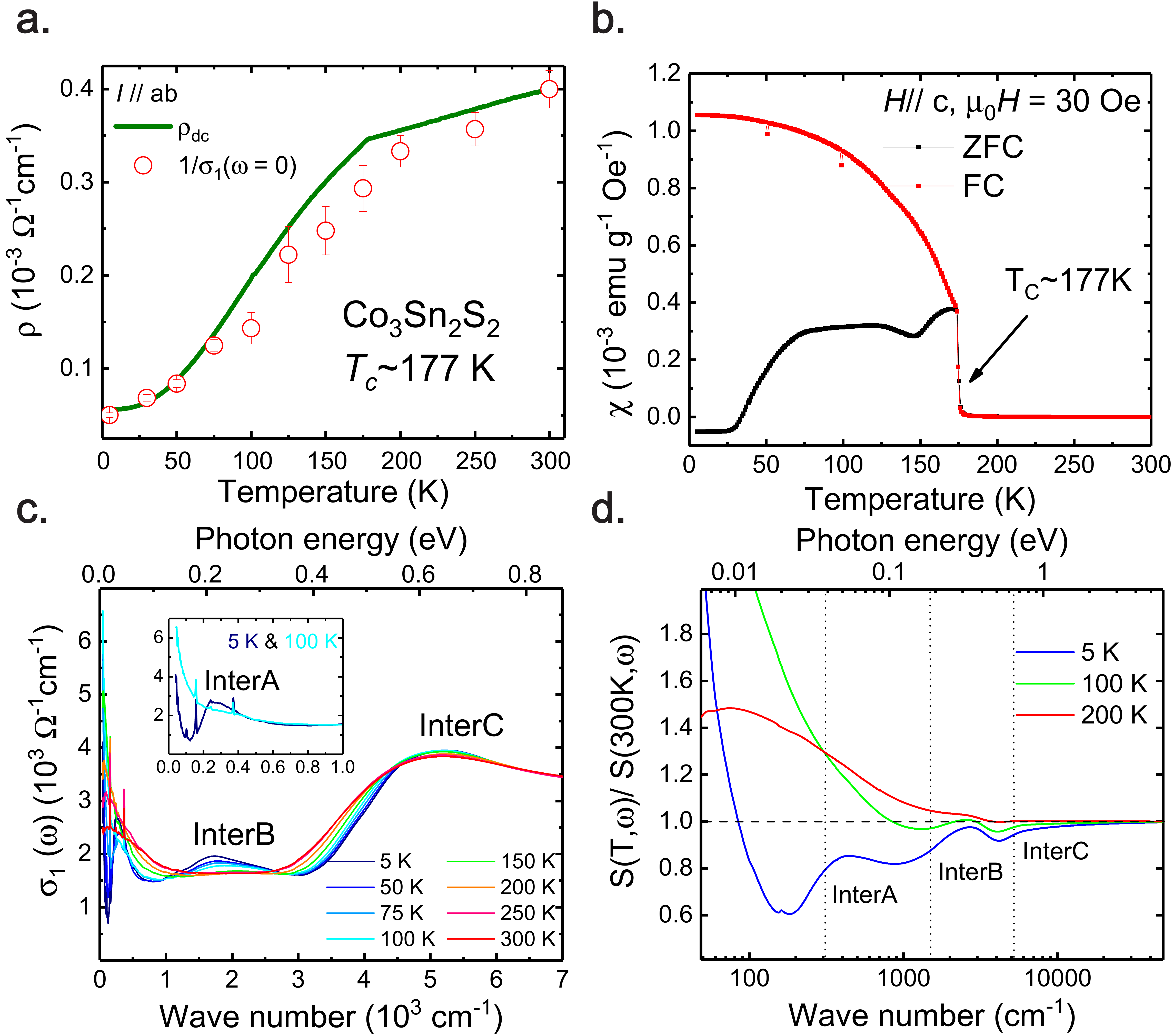}
}
\caption{(a)The dc resistivity ($\rho$) of Co$_{3}$Sn$_{2}$S$_{2}$ (solid line) with the zero-frequency values of optical conductivities(red circles). (b) The $T$ dependence of magnetic susceptibility $\chi(T)$ with zero-field-cooling and field-cooling mode at $\mu_{0}H$=30~Oe for $H||c$. The black arrow denotes the Curie Weiss temperature $T_{C}\sim$ 177~K. (c) The $T$-dependent optical conductivity $\sigma_{1}(\omega)$ from 30 to 7\,000~cm$^{-1}$. Inset shows the optical conductivity at 5 and 100~K in the far-infrared range. Three inter-band absorption peaks from low to high energy are denoted by `InterA',`InterB' and `InterC', respectively. (d) The ratio of the integrated spectral weight $S(T, \omega)/S(300~K, \omega)$ as a function of cutoff frequency ($\omega$) at different temperatures. $S(T, \omega)=\frac{Z_{0}}{\pi^2}\int^{\omega}_{0}~\sigma_{1}(\omega^{\prime}, T)d\omega^{\prime}$, in which $Z_{0}=$377$~\Omega^{-1} $is the vacuum impedance.}
\label{fig:sigma}
\end{figure}
The $T$ dependence of the real part of the in-plane optical conductivity [$\sigma_{1}(\omega)$] is shown in the energy range from 20~cm$^{-1}$ (2.5~meV) to 7\,000~cm$^{-1}$ (0.87~eV) (Fig. ~\ref{fig:sigma}c, see details of the measurements in the supplementary material~\cite{SM}).
The conductivities at 100~K (blue) and 5~K (black) are shown in the infrared region for comparison (inset of Fig.~\ref{fig:sigma}c).
At first glance, they consist of a Drude peak at zero-frequency and several other Lorentz peaks at finite frequencies.
Besides, several sharp features in conductivities are observed at 155~cm$^{-1}$, 245~cm$^{-1}$ and 370~cm$^{-1}$, which comes from the infrared-active vibrations.
Based on Kubo function, $\sigma_{1}(\omega)$ is proportional to the joint density of states, the peak at zero frequency (Drude) represents the intra-band response, and the one at a finite frequency (Lorentz) comes from the inter-band transition~\cite{Dressel2002}.
The extrapolated values for the \emph{dc} conductivity [$\sigma_{1}(\omega\rightarrow 0)\equiv~\sigma_{dc}$], circles in Fig.~\ref{fig:sigma}a are virtually identical to the resistivity, indicating excellent agreement between optical and transport measurements.
Upon cooling, $\sigma_{1}(\omega)$ shows apparent $T$ dependence, especially in the mid-infrared range, indicating that the band structure varies prominently with decreasing $T$.

At room temperature (300~K), the optical conductivity shows a typical metallic character, with a Drude-like free-carrier response at zero frequency.
Besides a Lorentz-like feature centered around 5\,000~cm$^{-1}$ and some phonon peaks, in the range from 1\,000 to 3\,000~cm$^{-1}$ the optical conductivity is almost frequency independent.
This plateau persists down to $T_{C}\sim$177~K, below which it gradually evolves into a Lorentzian, which grows up and moves to lower energy range with decreasing $T$.
Meanwhile, the peak centered at 5\,000~cm$^{-1}$ starts to strengthen and moves to higher energy.
These features indicate that, in the FM state, the band structure changes dramatically with $T$.
The Drude peak at zero frequency narrows continuously with decreasing $T$, implying suppressed quasiparticle scattering.
Below 100~K, the intra-band response is greatly suppressed, evolving into two components: an ultra-sharp Drude peak and a new interband absorption peak at 310~cm$^{-1}$(inset of Fig.~\ref{fig:sigma}c).
In the corresponding reflectivity (Fig. S1 in the supplementary material~\cite{SM}), a sharp plasma edge emerges below 100~K, indicating much coherent quasiparticle response~\cite{Sandilands2017, Yang2017}.

From the spectral weight of $\sigma_{1}(\omega)$(Fig.~\ref{fig:sigma}d), we see that, from 300 to 200~K, the $S(T, \omega)$ is gradually transferred from high to low energy range, indicating an enhanced intra-band response~\cite{Basov2005}.
While, below $T_{C}\sim$ 177~K , spectral weight below 1\,000~cm$^{-1}$ is transferred back to high energy range, accumulating in peaks at 2\,000 and 5\,000~cm$^{-1}$, respectively.
Correspondingly, in Fig.~\ref{fig:sigma}d, upon cooling, the spectral weight at the far-infrared range is gradually transferred to the mid-infrared absorption peaks (i.e., interB and interC).
However, below 100~K, the spectral weight of low-energy intra-band response is greatly suppressed, giving rise to a new absorption peak around 310~cm$^{-1}$ (38~meV, interA in the inset of Fig. ~\ref{fig:sigma}c).
Although such behavior is similar to the spin/charge-density-wave transition~\cite{Dai2016, Hu2008}, during which a gap emerges on the Fermi surface, in Co$_{3}$Sn$_{2}$S$_{2}$, no phase transition was predicted and observed in previous studies~\cite{Wang2018, Liu2018}.

%
%
\begin{figure}[tb]
\centerline{
\includegraphics[width=1\columnwidth]{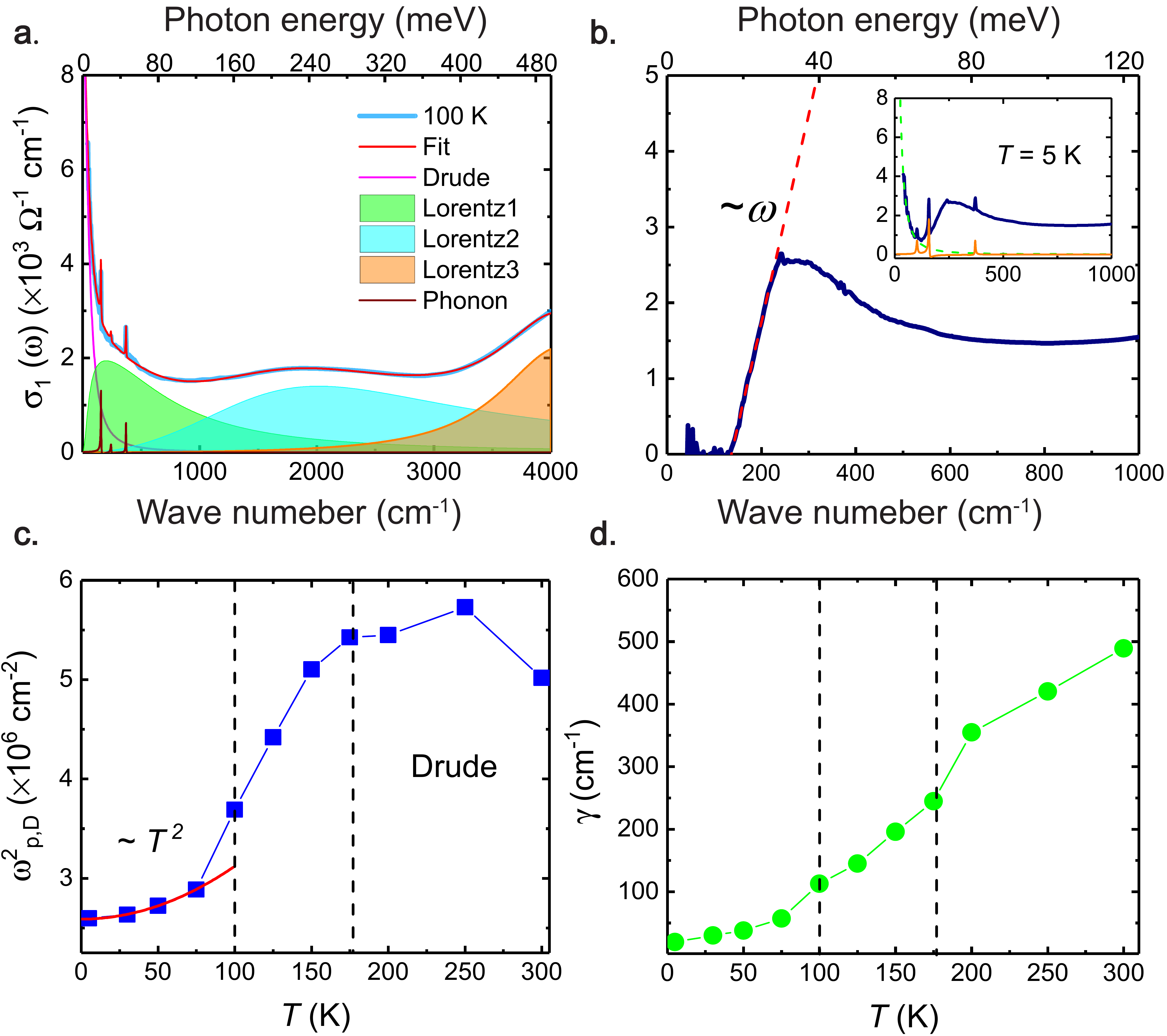}
}
\caption{(a)The Drude-Lorentz fit to the complex optical conductivity of Co$_{3}$Sn$_{2}$S$_{2}$ at 100~K. (b)The optical conductivity at 5~K, after the Drude peak and phonon peaks below 300~cm$^{-1}$ have been subtracted. Inset: the whole optical conductivity at 5~K, the green dashed line represents the intra-band response (Drude), origin line is the fit to the phonon modes below 300~cm$^{-1}$ (with a series of Lorentz peaks). (c) and (d) are the fitted value for the square of plasma frequency($\omega_{p,D}^{2}$) and the scattering rate($\gamma$) of Drude component, the red line in (c)is the $T^{2}$ fit to the square of plasma frequency. The dashed lines in (c) and (d) denotes the Curie Temperature ($T_{C}\sim~177~K$) and 100~K.  }
\label{fig:sigma fit}
\end{figure}
To get some more quantitative estimates of parameters determining the optical response, we fit the optical conductivity using a Drude-Lorentz model with the dielectric function $\tilde{\varepsilon}=\varepsilon_{1}+i\varepsilon_{2}$~\cite{Yang2015, Dressel2002}
%
%
\begin{equation}
\label{DrudeLorentz}
 \epsilon(\omega)=\epsilon_{\infty}-\sum_{i}\frac{\omega^{2}_{p,D;i}}
 {\omega^{2}+\frac{i\omega}{\tau_{D,i}}}+\sum_{j}\frac{\Omega^{2}_{j}}{\omega^2_{j}-\omega^2-i\omega\gamma_{k}},
\end{equation}
in which $\epsilon_{\infty}$ is the real part at high frequency.
In the first sum $\omega^{2}_{p,D;i}= 4\pi n_{i}e^{2}/m^{*}_{i}$ and $1/\tau_{D,i}$ are the square of plasma frequency and scattering rate for the delocalized (Drude) carriers, respectively, and $n_{i}$ and $m^{*}_{i}$ are the carrier concentration and effective mass.
In the second summation, $\omega_{j}$, $\gamma_{k}$ and $\Omega_{k}$ are the position, width and strength of the $j$th vibration or bound excitation.
The complex conductivity is $\tilde{\sigma}(\omega)=\sigma_{1}+i\sigma_{2} =-2\pi i\omega[\tilde{\varepsilon}(\omega)-\varepsilon_{\infty}]/Z_{0}$ ($Z_{0}\simeq 377~\Omega$ is the impedance of free space), and the real part $\sigma_{1}(\omega)$ is fitted using a non-linear least-squares technique.

The fit to the data at 100~K shown in Fig.~\ref{fig:sigma fit}a indicates that the optical conductivity can be reproduced quite well using one Drude peak and three Lorentz oscillators at 310, 1600 and 5\,100 cm$^{-1}$ ($\simeq$ 38, 198 and 632~meV).
Below 100~K, Drude response is suppressed dramatically, revealing the absorption peak centered at 310 cm$^{-1}$ (Inset of Fig.~\ref{fig:sigma}c).
We realize that it's hard to describe this peak with a simple Lorentz peak for its steep edge on the low-energy side.
To investigate this low-energy interband transition, in Fig. 2b, we have subtracted the Drude response and sharp phonon peaks (inset of Fig. 2b)~\cite{Xu2018a}.
In the residual conductivity, which purely comes from interband transitions, an unusual feature is a $\omega$-linear conductivity from 130 to 230~cm$^{-1}$ (16-29 meV), which indicates the presence of 3D linear bands near the Fermi level~\cite{Xu2016, Xu2018a, Timusk2013, Armitage2018}(see more discussion in Supplementary material~\cite{SM}).

The $T$ dependence of the square of plasma frequency ($\omega_{p,D}^{2}$), i.e., the Drude weight, and scattering rate ($1/\tau_{D}$) of Drude response are shown in Figs.~\ref{fig:sigma fit}c and d, respectively.
As the temperature is reduced, the Drude weight is firstly enhanced a little at 200~K and then suppressed after the FM transition.
Below 100~K, $\omega_{p,D}^{2}$ is considerably suppressed and shows $T^2$ dependence, which agrees well with theoretical predictions for Weyl semimetals~\cite{Burkov2011}.
The $1/\tau_{D}$ decreases dramatically with the $T$, from 495~cm$^{-1}$ at 300~K to 15~cm$^{-1}$ at 5~K.
Here, we notice that even though previous investigations on Co$_{3}$Sn$_{2}$S$_{2}$ predict the emergence of Weyl points right below $T_{C}\sim177~K$~\cite{Xu2018}, we only observed the distinct signature of linear bands at $T<100~K$.
The optical response indicates that, above 100~K, the response from other parabolic bands may superimpose over the response from the bands with linear dispersion.
Below 100~K, the parabolic bands may gradually move away from Fermi level, resulting in a Lifshitz transition, which leaves the linear bands dominating the low-energy response.
At 5~K, both the steep plasma edge in reflectivity (Fig. S1 in supplementary material~\cite{SM}) and the ultra-narrow Drude peak in optical conductivity reflect the coherence of the linear bands~\cite{Shao2019, Xu2016}.

\begin{figure*}[tb]
\centerline{
\includegraphics[width=2\columnwidth]{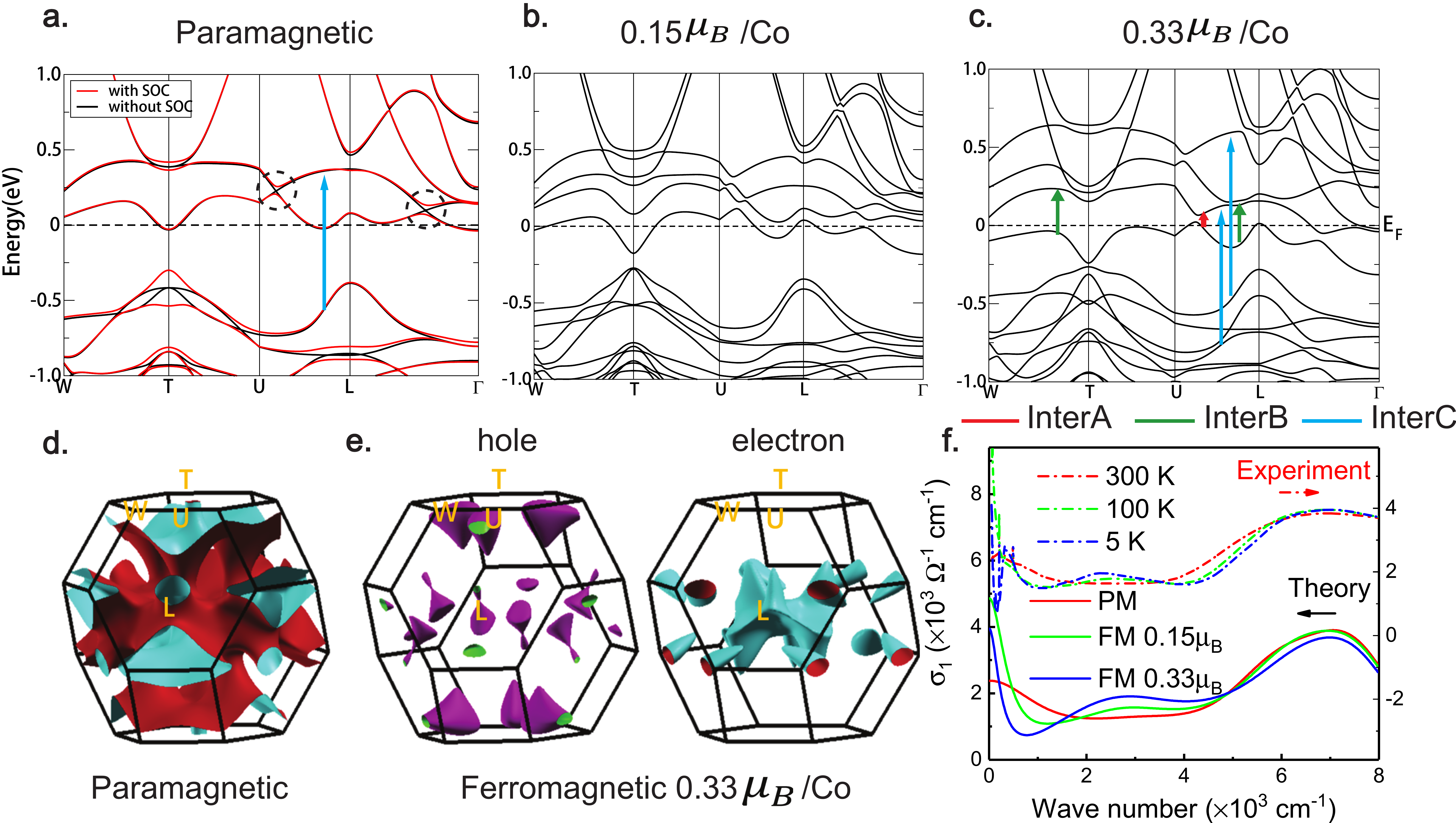}
}
\caption{(a) The band structure along high-symmetry paths with (red) and without (black) spin-orbital coupling (SOC). The dashed circles denote the band crossing opened by SOC. (b) and (c) show the band [$0.15\mu_{B}$/Co (b), $0.33\mu_{B}$/Co (c)]. Comparing with the measured optical conductivity, we ascribe the three absorption peaks in Fig. ~\ref{fig:sigma}c to inter-band transitions `interA', `InterB' and `InterC', respectively. The band structures in (b) and (c) are calculated with SOC. (d) and (e) are Fermi surfaces in paramagnetic (PM) and ferromagnetic (FM, $0.33\mu_{B}$/Co) state. (f) displays the simulated optical conductivities (solid lines) based on band structure calculation (with SOC) in the PM and FM ($0.15\mu_{B}$/Co and $0.33\mu_{B}$/Co) states in comparison with experimental results (dashed dot lines). The energy discrepancy between the experimental and theoretical results mainly comes from the correlation effect, which is not considered during the calculation.}
\label{fig:bands}
\end{figure*}
%

To better understand the optical conductivity as well as the electronic properties, we have simulated the band structure and optical conductivity of Co$_{3}$Sn$_{2}$S$_{2}$ through the first-principle calculations.
Because of the weaker magnetic fluctuation as $T$ drops, the effective moment on Co atoms increases upon cooling.
To mimic the $T$ effect, the magnetic moment on Co is constrained at different values during the self-consistent calculation.
In paramagnetic (PM) state, the calculated band structure shows semi-metallic feature in Fig.~\ref{fig:bands}(a).
Without including SOC, there exist two linear band crossings marked in dashed circles in $U-L$ and $L-\Gamma$ paths as protected by the mirror symmetry.
These crossing points are on the nodal rings in the mirror planes~\cite{Wang2018}.
On considering SOC, band gap opens along the rings.
Fig.~\ref{fig:bands}(b) and (c) are band structures in the cases of 0.15 $\mu_B$ and 0.33 $\mu_B$ on Co atoms, respectively.
The later one is in the ground state without any constraint.
The spin exchange splitting gradually separates the bands into spin up and spin down channels. The spin-up one shifts closer to the Fermi level, and the spin-down moves to higher energy. In the case without SOC (Fig. S5), the nodal rings protected by mirror plane are kept well in both channels.
The Fermi surfaces in PM and FM are different as shown in Figs.~\ref{fig:bands} (d) and (e). It becomes a half-metal in the ground state (Fig.~S5b in the supplementary material~\cite{SM}).
When SOC is further considered, the crossing points open gaps along the nodal rings and away from the mirror planes, three pairs of Weyl points appears. The band structure around Weyl point is described in Fig. S2~\cite{SM} and Ref.~\cite{Liu2018, Wang2018, Xu2018}.

Next, we calculated the optical conductivity with different magnetization.
In addition to the inter-band contributions, the intra-band transition is included as the Drude model.
The plasma frequency $\omega_{p}$ contributed to the Drude peak can be obtained from band structure calculations.
It is obtained as 1.73 eV for PM and 1.34 and 0.94 eV for FM with 0.15 and 0.33 $\mu_B$/Co, respectively.
$\omega_{p}$ decreases with magnetization, which is consistent with the shrinking of Fermi surfaces in Figs~\ref{fig:bands}(d) and (e), as well as the decreasing of carrier density~\cite{Liu2018}.
The scattering rate $\frac{1}{\tau_{D}}$ is parameterized as 0.17 eV, 0.05 eV and 0.03 eV to mimic the decreasing (increasing) of $T$ (lifetime) and to well fit the experimental spectra.
Within these settings, the theoretical spectra are shown in Fig.~\ref{fig:bands}(f) together with the experimental ones.
It is noted that the experimental spectra are replotted with the incident photon wavenumber being scaled by a factor of 1.33 to match the InterC in Fig.~\ref{fig:sigma}c with the calculated position. This difference between the experimental and theoretical results comes from the correlation effect due to 3$d$ orbitals of Co, which is not considered during the calculation within local density approximation. The overall consistency between the calculated and scaled experimental spectra indicates that the single-particle approximation can catch the most essential features of this material and the electron correlation is moderate in Co$_3$Sn$_2$S$_2$~\cite{Qazilbash2009}.

One can also notice that the optical conductivity around InterC is not sensitive to the magnetization or spin splitting in bands, while InterB is very sensitive and appears after magnetic phase transition.
Considering the optical selection rule, including inter-transition energy, the enhanced joint density of states from roughly parallel bands and the angular momentum change without spin flipping, we attribute the InterC coming from to the interband transitions marked by blue arrows in Fig.~\ref{fig:bands}a and c.
The associated occupied and unoccupied bands are from Co 3$d$ $t_{2g}$ and $e_g$ orbitals, repspectively.
These bands are separated due to octahedral crystal-field splitting being irrelevant to the spin-exchange splitting.
For the InterB, the interband contributions are marked by green arrows, and they are mainly the transitions among bands of Co $3d$ orbitals in nearly the same spin channel, which is reasonable since there is nearly non spin flipping and sensitive to magnetization.

In the experimental data, below 100~K, besides InterB and InterC, we observed another absorption peak around 310~cm$^{-1}$ (InterA in Fig.~\ref{fig:sigma}c).
However, this peak cannot be resolved in calculated spectra (Fig. ~\ref{fig:bands}f). This may come from the overlapping of the low-energy peak with the overestimated Drude response.
To resolve this, we established a the tight-binding Hamiltonian from Wannier function and performed quite dense k-sampling for integration of inter-band transitions.
A small absorption peak can be seen after SOC is considered, which is indicated by the red arrow in Fig. S4 in the supplementary material~\cite{SM}. This peak is around 350~cm$^{-1}$, which is consistent with the position of InterA in Fig.~\ref{fig:sigma}c. Checking the bands near the Fermi level, we notice that the SOC only triggers gaps along the nodal line except for isolated Weyl points. Thus, we ascribe the InterA to the interband transitions between the inverted bands which are gapped by SOC as marked by red arrow in Fig.~\ref{fig:bands}c. These bands are very close to and compose of Weyl nodes.

In summary, we have carried out comprehensive optical and theoretical investigations on the electronic properties of a magnetic Weyl semimetal candidate Co$_{3}$Sn$_{2}$S$_{2}$.
Experimental and theoretical results reveal a moderate electron correlation and show that the increasing exchange splitting renormalize the bands continuously, making the SOC induced gap approaching Fermi level. We have clearly identified the bands contributing to each absorption peak, including the magnetization sensitive and nonsensitive ones. The one from bands close to Weyl nodes is also identified. These results strongly suggest that Weyl nodes can be well controlled by magnetization, which is quite promising for potential applications.

%
%
%
We thank Q. Niu, S. Q. Shen, Y. F. Xu and Z. Y. Qiu for useful discussions.
X. G. Q acknowledges the support from NSFC (Project No. 11374345 and No. 91421304) and  MOST (Project No. 2015CB921303 and No. 2015CB921102).
H. M. W acknowledges the support from the Ministry of Science and Technology of China under grant numbers 2016YFA0300600 and 2018YFA0305700; the National Science Foundation of China under grant numbers 11674369; the K. C. Wong Education Foundation (GJTD-2018-01); Beijing Municipal Science \& Technilogy Commission (Z181100004218001) and Beijing Natural Science Foundation (Z180008).
%
%
%
%
%
%

%

\end{document}